\documentstyle[11pt]{article}
\begin{document}
\begin{titlepage}
\vspace{15mm}
\begin{center}
{\Large {\bf Motion of test particles in six-dimensional dilatonic
Kaluza-Klein theory} }
\vspace{6mm} \\
{\bf M. Biesiada}${}^{\dag}$ \footnote{mb@imp.sosnowiec.pl},
{\bf J. Syska}${}^{\ddag}$ \footnote{jacek@server.phys.us.edu.pl}\\
\vspace{5mm}
${}^{\dag}$
{\sl Institute of Physics,}\\
{\sl University of Silesia, Uniwersytecka 4, 40-007 Katowice, Poland}\\
{\sl Department of Astrophysics and Cosmology,}\\
\noindent ${}^{\ddag}$
{\sl Modelling Research Institute, Drzyma{\l}y 7 m.5, 40-059 Katowice, Poland}\\
{\sl and Department of Field Theory and Particle Physics, Institute of Physics,} \\
{\sl University of Silesia, Uniwersytecka 4, 40-007 Katowice, Poland}\\
\end{center}

\setcounter{equation}{0} \vspace{5 mm}

\begin{abstract}

Multidimensional theories still remain attractive from the point
of view of better understanding of fundamental interactions.
In this paper we consider a six - dimensional Kaluza -- Klein
type model at the classical level. We derive static spherically
symmetric solutions to the multidimensional Einstein equations.
They are fundamentally different from four - dimensional
Schwarzschild solutions: they are horizon free and the presence
of massless dilaton field has the same dynamical effect as the
existence of additional
massive matter in the system. \\
Then we analyse the motion of test particles in such spherically symmetric configurations. The emphasis is put on some observable quantities like redshifts.  It is suggested that strange features of emission lines from active galactic nuclei as well as quasar - galaxy associations may in fact be manifestations of multidimensionality of our world. \\

\vspace{20 mm}

Kaluza -- Klein type models, Hamilton-Jacobi equation, redshifts,
quasar - galaxy associations

\end{abstract}

\vspace{35mm}


\vspace{2mm} \vfill \vspace{5mm}
\end{titlepage}

\section{Intoduction}
\label{noncosm-intr}

\hfill{And do not be called teachers;

\hfill{for One is your Teacher,}

\hfill{the Christ {\it Jesus}.}

\vspace{2 mm}

\hfill{Holy Bible, Matthew 23,10}

\vspace{7 mm}

Many recent ideas in theoretical physics assume the possibility that our world may have more than four dimensions. The discovery of the cancellation of infinities in superstring theories has stimulated the interest in higher-dimensional theories and their compactification.
The renewed interest in Kaluza-Klein theories stems from the fact that multidimensional analogues of general relativity are able to generate non-Abelian gauge theories out of theories with symmetries of compact
internal space.
Multidimensional versions of supersymmetric theories are examples of another approach to construct a theory of elementary interactions \cite{Gr-Sch-1981},\cite{Gr-Sch-1982b},\cite{Kaku-1988} by exploiting the formal symmetry between bosons and fermions. These theories could provide better description of observational aspects of the world we see. However, the main problem of multidimensional models is their complexity reflected either in  analytical or numerical studies --- even in the case of models with the Ricci flat Calabi-Yau manifolds.

Up to now there is no well established experimental evidence of six-dimensionality of our world and our understanding of potential manifestations of higher dimensions is too poor. But there were also attempts in the literature to seek the effects of extra dimensions in the astrophysical setting pursued actively by
Wesson \cite{Wesson-1992}, Lim, Kalligas, Everitt, Biesiada \cite{LORD-Biesiada-Syska-Manka,Biesiada-Rudnicki-Syska-1b}, Ma\'{n}ka, and Syska. This line of thinking is worth of developing in order to gain better understanding concerning possible manifestations of six-dimensionality of the world. In particular this would mean that the effects of extra dimensions may be well around us on the contrary of standard expectations that extremely high energies are necessary to probe the higher dimensions.

Hence the contents of the present paper suggest that there is an intimate ralation between the manifestations of six-dimensionality of this world and the so-called dark matter. But the first evidence for the existence of the dark matter in galaxies came into play when Oort in 1932 \cite{Oort-32} and Zwicly in 1933 have applied the famous virial theorem to vertical motion of stars in the Galaxy and to the radial velocities of members of Coma cluster, respectively. The problem revived and become well established in the seventies when it has been demonstrated \cite{Rubin-Ford-70,Rubin-Ford-Thonnard-80,Gates-Gyuk-Turner} that the rotation curves of spiral galaxies were indicative to the presence of unseen\footnote{\label{f-cyt-6}
The word {\it unseen} would perhapse reflect better the nature of the missing mass being free from potential negative connotations associated with the word {\it dark}. However in the rest of this work we shall use the traditional notion of {\it dark matter}.} (in any part of electromagnetic spectrum) mass\footnote{\label{f-cyt-7}
There are currently many candidates for the dark matter with massive compact halo objects (MACHOs) such like "brown dwarfs", "jupiters" etc. on one extreme to hypothetical elementary particles such like massive neutrinos, axions and other weakly interacting massive particles (WIMPs) which till now are either unable to give the amount of needed mass or are in contradiction with other parts of proposed models.}. In their paper \cite{Bahcall-Lubin-Dorman} Neta Bahcall, Lubin and Dorman point to the evidence that most dark matter in the universe resides in large dark halos around galaxies.

Thus the present paper provides a description of properties of certain six-dimensional Kaluza-Klein type model with pointing to some of possible observational consequences.
Section~2 contains a description of the model along with motivations for the choice of six-dimensional spacetime. Then we derive static spherically symmetric solutions of the multidimensional Einstein equations \cite{LORD-Biesiada-Syska-Manka}. They are in a sense analogous to the familiar four-dimensional Schwarzschild solution but fundamentally different i.e. they are horizon free. More detailed discussion of the properties of these solutions is presented in Sections~2 and  3. We also discuss briefly some strictly observable quantities in the model such like the redshift formulae.
Section~4 is devoted to the analysis of motion of test particles.
Finally Section~5 contains concluding remarks and perspectives. Some more formal although important remarks are put into Appendices in the end of this paper. Throughout Section~2 we are using natural units (c = $\hbar$ =1) whereas in
Sections~3,~4~and~5 which deal with some more observationally
oriented issues we reintroduce explicitly the velocity of light (the Planck constant turns out to be irrelevant for these considerations).

\section{Field equations}
\label{general}

Let us consider a six-dimensional field theory comprising the gravitational self field described by a metric tensor, $g_{MN}$, and a real massless "basic" scalar field, $\varphi$. The scalar field $\varphi$ is a dilaton field--hence, just below, the minus sign in front of its kinetic energy term
\cite{LORD-Biesiada-Syska-Manka,Biesiada-Rudnicki-Syska-1b}.
In a standard manner we decompose the action into two parts:
\begin{equation}  \label{row_dzialanie-EH-fi}
{\cal S} = {\cal S}_{EH} + {\cal S}_{\varphi} \; ,
\end{equation}
where $S_{EH}$ is the Einstein --- Hilbert action
\begin{equation}  \label{row_1}
{\cal S}_{EH} = \int \frac{1}{2 \kappa_{6}} \: \sqrt{- g} \: {\cal R} \: d^{6}x
\end{equation}
and $S_{\varphi}$ is the action for a real massless scalar field\footnote{\label{f-cyt-created-fluct}
This scalar field $\varphi$ is the created fluctuation out from nothingness.
}
\begin{eqnarray}  \label{row_2}
{\cal S}_{\varphi} &=& - \int \sqrt{- g} \: \frac{1}{2} \: g_{MN} \:
\partial^{M} \varphi \partial^{N} \varphi \: d^{6}x =  \nonumber \\
&=& \int \sqrt{- g} \: {\cal L}_{\varphi} \: d^{6} x \; .
\end{eqnarray}
In Eq.(\ref{row_2}), $g = det\; g_{MN}$ denotes the determinant of the metric tensor, ${\cal R}$ is the curvature scalar of six - dimensional (in general curved) spacetime, and $\kappa_{6}$ denotes the coupling constant of the six-dimensional theory, analogous to familiar Newtonian gravity constant.
${\cal L}_{\varphi}$ is the Lagrangian density for a scalar massless field~$\varphi$.

The motivations for choosing by many others six-dimensional models were diverse. Nishino, Sezgin \cite{Nishino-Sezgin}, Salam and Sezgin \cite{Salam-Sezgin-1984} have suggested that one may obtain
the fermion spectrum in four-dimensions within the framework of D=6, N=2 Kaluza-Klein supergravity.
Their Lagrangian density was equal to
\[
{\cal L} = {\cal L}_{EH} + {\cal L}_{m} \; ,
\]
where:
\begin{eqnarray}
{\cal L}_{EH} &=& \frac{1}{4\;\kappa^2}\: {\cal R}  \nonumber \\
\!\!\!\!\!{\cal L}_{m} &=& -\frac{1}{2}\:g^{MN}\partial_M\varphi\:\partial_N\varphi - \nonumber \\
&-& \frac{1}{4} \kappa\exp{(\sqrt{2}\kappa \varphi)}\;F_{PQ}F^{PQ} -  \nonumber \\
&-& \frac{1}{2} \left(g_0^3/\kappa^3\right)\:\exp{(-\sqrt{2}\kappa \varphi)}
\; , \nonumber
\end{eqnarray}
In the above equation, $F_{PQ} = \partial_P A_Q - \partial_Q A_P$, $\varphi$, and $A_P \:(P=0,...,6)$ are boson fields, $g_0$ is the U(1) coupling constant. The mathematical compactification of the full six-dimensional space into a direct product $M^4 \times S^2$ is realized with the ground state expectation values:
$\varphi = \varphi_0$, and $F_{MN} = F_{mn} = -\frac{1}{2}g_0b^2 \varepsilon_{mn}$ for $M,N = m,n$, and $0$ otherwise ($b$ is the radius of the microspace). Our model is similar to the above model (considered in Refs.\cite{Nishino-Sezgin} and \cite{Salam-Sezgin-1984}), in absence of boson fields\footnote{The inclusion of gauge fields into the model and its consequences for the problem of dark matter will be a subject of forthcoming papers.
}.
A six-dimensional model of the Kaluza-Klein theory was also previously investigated by Ivashchuk and Melnikov \cite{Iv-Mel-1995,Iv-Mel-1994}, Bronnikov and Melnikov \cite{Br-Mel-1995}, and by one of us in \cite{Manka-Syska-a}.

The simplest extension of familiar four-dimensional spacetime is to consider five-dimensional models, as Wesson considered in \cite{Wesson-1992}.
We have chosen a six - dimensional theory which is much more
robust and interesting.

By extremalizing the action given by Eqs.(\ref{row_1})-(\ref{row_2}) we obtain the Einstein equations
\begin{equation}  \label{row_3}
R_{MN} - \frac{1}{2} \: g_{MN} \: {\cal R} = \kappa_{6} \: T_{MN}.
\end{equation}
Here $R_{MN}$ is the six - dimensional Ricci tensor, ${\cal R}$ is the
six-dimensional curvature scalar and $T^{M}_{N}$ is the energy - momentum
tensor of a real scalar field $\varphi$ which is given by
\begin{equation}  \label{row_4}
T^{M}_{N} = \partial_{N} \varphi \: \frac{\partial {\cal L}_{\varphi}}
{\partial (\partial_{M} \varphi )} - \delta^{M}_{N} {\cal L}_{\varphi} \; .
\end{equation}

Variation of the total action ${\cal S}$ with respect to the field $\varphi$
gives the Klein - Gordon equation
\begin{equation}  \label{row_5}
\Box \varphi = 0 \;\;  ,
\end{equation}
where
\[
\Box = - \frac{1}{\sqrt{- g}} \: \partial_{M} (\sqrt{-g}\; g^{MN}
\partial_{N})
\]
and $g^{MN}$ is the tensor dual to $g_{MN}$ .

Now, we assume that we live in the compactified (which is quite a reasonable
assumption) world, where the six - dimensional spacetime is a topological
product of ``our'' curved four - dimensional
physical spacetime (with the metric $g_{\alpha \omega }, \; \alpha , \omega
= 0,1,2,3 $) and the internal space (with the metric $g_{he}, \; h,e=5,6 $).

Therefore the metric tensor can be factorized as
\begin{eqnarray}  \label{row_6}
g_{MN} = \pmatrix{g_{\alpha \omega } & 0 \cr 0 & g_{he} } \; .
\end{eqnarray}
The four-dimensional diagonal part is assumed to be that of a spherically
symmetric geometry
\begin{eqnarray}  \label{row_7}
g_{\alpha \omega} = \pmatrix{e^{\nu (r)} & & & \cr & - e^{\mu (r)} & 0 & \cr
& 0 & - r^{2} & \cr & & & - r^{2} sin^{2}\Theta \cr } \; ,
\end{eqnarray}
where $\nu (r)$ and $\mu (r)$ are (at this stage) two arbitrary functions. \\
Analogously, we take the two-dimensional internal part to be
\begin{eqnarray}  \label{row_8}
g_{he} = \pmatrix{ - \varrho^{2} (r) \: cos^{2} \vartheta & 0 \cr 0 & -
\varrho^{2} (r) } \; .
\end{eqnarray}
The six-dimensional coordinates $(x^{M})$
are denoted by $(t, r,\Theta,\Phi ,\vartheta,\varsigma) $ where $t \in
[0,\infty )$ is the usual time coordinate, $r\in [0,\infty ),$ $\Theta \in
[0,\pi ] $ and $\Phi \in [0,2 \pi ) $ are familiar three-dimensional
spherical coordinates in the macroscopic space;
$\vartheta \in [-\pi,\pi)$
and $\varsigma \in [0, 2 \pi )$ are coordinates in the internal
two-dimensional space and $\varrho \in (0,\infty )$ is the ``radius'' of this
two-dimensional internal space. We assume that $\varrho (r) $ is the
function of the radius $r$ in our external three-dimensional
space\footnote{\label{f-cyt-9}
special thanks to Ryszard Ma\'{n}ka for pointing to this direction}.

The internal space is a 2-dimensional topological torus with
$r$-dependent parameter $\varrho (r)$ which can be represented as
a surface embedded in the three-dimen\-sio\-nal Euclidean space
\begin{eqnarray}  \label{row_9}
\left\{
\begin{array}{lll}
w^{1} = \varrho (r) \: cos \varsigma \;\; , \;\; \varsigma \in [0, 2 \: \pi )
&  &  \\
w^{2} = \varrho (r) \: sin \varsigma \;\; &  &  \\
w^{3} = \varrho (r) \: sin \vartheta \;\; , \;\; \vartheta \in [- \pi, \pi )
\; .  &  &
\end{array}
\right.
\end{eqnarray}
\nopagebreak[4]

Now using Eqs.(\ref{row_7})-(\ref{row_8}), we can calculate the
components of the Ricci tensor. The nonvanishing components are
\begin{eqnarray}  \label{row_10}
R^{t}_{t} &=& \left( 4 \: \varrho^{2} r \nu^{\prime}+ 4 \: r^2
\varrho^{\prime}\varrho \nu^{\prime} - r^{2} \varrho^{2}
\mu^{\prime}\nu^{\prime} + \right. \nonumber \\
&+& \left. r^{2} \varrho^{2} (\nu^{\prime})^{2} + 2 \: r^{2}
\varrho^{2} \nu^{\prime\prime} \right) (4 \: e^{\mu} r^{2} \varrho^{2})^{-1}
\end{eqnarray}
\begin{eqnarray}  \label{row_11}
R^{r}_{r} &=& \left(- 4 \: \varrho^{2} r \mu^{\prime} - 4 \: r^2
\varrho^{\prime}\varrho \mu^{\prime} - r^{2} \varrho^{2}
\mu^{\prime}\nu^{\prime}+  \right. \nonumber \\
&+& r^{2}  \varrho^{2} (\nu^{\prime})^{2} +
\left. 8 \: r^{2} \varrho \varrho^{\prime\prime} + \right. \nonumber \\
&+& \left. 2 \: r^{2} \varrho^{2} \nu^{\prime\prime} \right)
\left( 4 \: e^{\mu} r^{2} \varrho^{2} \right)^{-1}
\end{eqnarray}
\begin{eqnarray}  \label{row_12}
R^{\Theta}_{\Theta} = R^{\Phi}_{\Phi} &=& \left( - 4 \: e^{\mu} \varrho^{2} +
4 \: \varrho^{2} + 8 r \varrho \varrho^{\prime}-  \right. \nonumber \\
&-& 2 \: r \varrho^{2} \mu^{\prime} +\left. 2 \: r \varrho^{2} \nu^{\prime} \right)
\left( 4 \: e^{\mu} r^{2} \varrho^{2} \right)^{-1}
\end{eqnarray}
\begin{eqnarray}  \label{row_14}
R^{\vartheta}_{\vartheta} = R^{\varsigma}_{\varsigma} &=& \left( 8 \: \varrho r \varrho^{\prime} +
4 \: r^2 (\varrho^{\prime})^{2} - 2 \: r^{2} \varrho \varrho^{\prime}\mu^{\prime} + \right. \nonumber \\
&+& \left. 2 \: r^{2} \varrho \varrho^{\prime}\nu^{\prime} + 4 \: r^{2} \varrho \varrho^{\prime\prime} \right)
\left( 4 \: e^{\mu} r^{2} \varrho^{2} \right)^{-1}
\end{eqnarray}

Let us assume that we are looking for a solution of the Einstein equations
(see Eq.(\ref{row_3})) with the Ricci tensor given by
Eqs.(\ref{row_10})-(\ref{row_14}), with
$\nu (r) = \mu(r)$, and with the following boundary conditions
\begin{eqnarray}  \label{row_16}
\lim_{r\rightarrow \infty } \nu (r) = \lim_{r\rightarrow \infty } \mu (r) = 0
\; ,
\end{eqnarray}
\begin{eqnarray}  \label{row_17}
\lim_{r\rightarrow \infty } \varrho (r) = d = constant \neq 0 \; .
\end{eqnarray}
In other words, we are looking for the solution which at spatial infinity
reproduces the flat external four-dimensional Minkowski spacetime and static
internal space of ``radius'' $d$ which is of order of $10^{-33} \, m.$

Now we make an assumption that the scalar field $\varphi $ depends
neither on time $t$ nor on internal coordinates $\vartheta$ and $\varsigma$.
Now, because of assumed spherical symmetry of the physical spacetime,
it is natural to suppose that the scalar field $\varphi$ is the function of
the radius $r$ alone, $\varphi = \varphi (r)$. We impose also the boundary
condition for the scalar field $\varphi$
\begin{equation}  \label{row_18}
\lim_{r\rightarrow \infty } \varphi (r) = 0 \;
\end{equation}
which supplements boundary conditions (\ref{row_16}) and (\ref{row_17})
for the metric components.

By virtue of Eqs.(\ref{row_4}) and (\ref{row_2}) it is easy to see that
the only nonvanishing components of the energy-momentum tensor are
\begin{eqnarray}  \label{row_19}
- T^{r}_{r} = T^{t}_{t} = T^{\Theta}_{\Theta} = T^{\Phi}_{\Phi} &=&
T^{\vartheta}_{\vartheta} = T^{\varsigma}_{\varsigma} = \nonumber \\
&=& \frac{1}{2} \: g^{rr} \: (\partial_{r} \varphi )^{2} \; .
\end{eqnarray}

Consequently, it is easy to verify that the solution of the Einstein equations
(\ref{row_3}) is
\begin{eqnarray}  \label{row_20}
\nu (r) = \mu (r) = \ln \left( \frac{r}{r + A} \right)
\end{eqnarray}
\begin{eqnarray}  \label{row_21}
\varrho (r) =d \: \sqrt{\frac{r + A}{r}}
\end{eqnarray}
\begin{eqnarray}  \label{row_22}
\varphi (r) = \pm \sqrt{\frac{1}{2 \kappa_{6}}} \:\ln \left( \frac{r}{r + A}
\right) \; .
\end{eqnarray}
Hence we obtain that the only nonzero component of the Ricci tensor (see
Eqs.(\ref{row_10})-(\ref{row_14})) is $R^{r}_{r}$ which
reads\footnote{\label{f-cyt-10}
Putting Eqs.(\ref{row_25}),(\ref{row_19})~and~(\ref{row_24})
together we can notice a similarity between the equation
\begin{displaymath}
\label{screen-in-gravity}
R^{r}_{r} = -\kappa_{6} (\partial_{r} \varphi)^{2} g^{rr} \;\;\;\;\;\;\;\; (*)
\end{displaymath}
and its electromagnetic analog
\begin{displaymath}
\nabla^2 {\bf A} = m_{A}^{2} {\bf A} \; ,
\end{displaymath}
where ${\bf A}$ is the electromagnetic vector potential.

Eg.($*$) is the screening current condition in
gravitation, analogous to that in electromagnetism.
}
\begin{eqnarray}  \label{row_23}
R^{r}_{r} = \frac{A^{2}}{2 \: r^{3} (r + A)} \; .
\end{eqnarray}
So the curvature scalar ${\cal R}$ is equal to
\begin{eqnarray}  \label{row_24}
{\cal R} = R^{r}_{r} = \frac{A^{2}}{2 \: r^{3} (r + A)} \; ,
\end{eqnarray}
where $A$ is the real constant, with the dimensionality of length, which value
is to be taken from observation for each particular system.

In derivation of the above solutions, we have used
Eqs.(\ref{row_10})-(\ref{row_14}) which together with
Eq.(\ref{row_19}), imply that all of the six diagonal Einstein equations are
equal to just one
\begin{eqnarray}  \label{row_25}
\frac{1}{2} \: {\cal R} = \kappa_{6} \: T^{r}_{r} \; .
\end{eqnarray}

Now we can rewrite the metric tensor in the form
\begin{eqnarray}  \label{row_26}
g_{MN} &=& diag \, (\frac{r}{r + A},\, - \frac{r}{r + A},\,
 - r^{2},\, - r^{2} sin^{2}\Theta,\, \nonumber \\
&&  -  d^{2} \frac{r + A}{r} cos^{2} \vartheta,\, - d^{2} \frac{r + A}{r} )
\end{eqnarray}
with its determinant equal to
\begin{eqnarray}  \label{row_27}
g = det g_{MN} = - (d^{2} \: r^{2} \: sin\Theta \: cos\vartheta )^{2} \; .
\end{eqnarray}
Thus, we see that the spacetime of our model is stationary.

It is also necessary to verify, whether the solution of the Klein-Gordon
equation (see Eq.(\ref{row_5})) is in agreement with Eq.(\ref{row_22}),
which follows from the Einstein equations. From Eq.(\ref{row_5}), we obtain
that
\begin{eqnarray}  \label{row_28}
\partial_{r} \varphi (r) = - C \: g_{rr} \: r^{-2} = C \: \frac{1}{r(r + A)}
\; ,
\end{eqnarray}
where C is a constant. Comparing this result with Eq.(\ref{row_22}) we
conclude that if
\begin{equation}  \label{row_29}
C = \pm \, \frac{A}{\sqrt{2 \kappa_{6}}}
\end{equation}
then the solution of the Klein-Gordon equation is in agreement with the
solution of the Einstein equations coupled to Klein-Gordon equation.
Hence, the real massless "basic" free
scalar field $\varphi (r)$ (see Eq.(\ref{row_22})) can be the source of the
nonzero metric tensor as in Eq.(\ref{row_26}). Only when the constant $A$ is
equal to zero, the solutions (\ref{row_20}) -- (\ref{row_22}) become trivial
and the six-dimensional spacetime is Ricci flat.

It is worth noting that because the components $R^{\vartheta}_{\vartheta}$
and $R^{\varsigma}_{\varsigma}$ of the Ricci tensor are equal to zero for
all values of $A$, the internal space is always Ricci flat. However, we
must not neglect the internal space because its ``radius'' $\varrho$ is a
function of $r$ and the two spaces, external and internal, are therefore
``coupled''. Only when $A = 0$ are these two spaces ``decoupled'', and our
four-dimensional spacetime becomes Minkowski flat\footnote{
When $A$ is not equal to
zero our four-dimensional external spacetime is curved and its scalar
curvature ${\cal R}_{4}$ could be given by Eq.(\ref{row_24})
\begin{eqnarray}  \label{row_30}
{\cal R}_{4} = {\cal R} = \frac{A^{2}}{2 \: r^{3} (r + A)} \; .
\end{eqnarray}
}.

\section{Some properties of solutions}
\label{prop}

If the parameter $A$ is strictly positive\footnote{
The discussion of $A<0$ case will be left for the Appendix~A.1.},
$A > 0,$ then Eqs.(\ref{row_20})-(\ref{row_22}) are valid for all $r > 0$. The
metric tensor becomes singular only at $r = 0$, nevertheless its determinant
$g$ (see Eq.(\ref{row_27})) remains well defined. Below, we shall collect
several formulae which will be useful in our later discussion. We start
with time and radial components of the metric $g_{MN}$ and the internal
``radius'' $\varrho (r)$ (see Eqs.(\ref{row_21})~and~(\ref{row_26}))
\begin{eqnarray}  \label{row_31}
\left\{
\begin{array}{lll}
g_{tt} = \frac{r}{r + A} \; &  &  \\
g_{rr} = - \frac{r}{r + A} \; &  &  \\
\varrho (r) = d \sqrt{\frac{r + A}{r}} \; \; . &  &
\end{array}
\right.
\end{eqnarray}
It will also be useful to write the explicit relation for the real, physical
radial distance $r_{l}$ from the center
\begin{eqnarray}  \label{row_32}
\!\!\!\!\! r_{l} & = & \int_{0}^{r} dr \, \sqrt{- g_{rr}} = {\sqrt{{\frac{r}{{r + A}}}}}\,\left( r + A \right) + \nonumber \\
&+& {\frac{1}{2}
\,A\,\ln ({\frac{A}{{A + 2\,r + 2\, (r + A) \,{\sqrt{{\frac{r}{{r + A}}}}} }}
})} < r \; .
\end{eqnarray}

Because $g_{tt} \rightarrow 1 $ for $r \rightarrow \infty $ (see
Eq.(\ref{row_31})), it is interesting to compare the gravitational potential
$g_{tt} = \frac{r}{r + A} \approx 1 - \frac{A}{r}$ for $r\gg A$ with the
gravitational potential $g_{tt} = 1 - \frac{G}{c^{2}} \frac{2 M}{r} $
induced by a mass $M$ in the Newtonian limit. $G$ and $c$ are the
four-dimensional gravitational constant and the velocity of light,
respectively. Comparing these two potentials, we obtain that
$A = 2  \frac{G}{c^{2} } M $, so the parameter $A$ can have the same
dynamical consequences\footnote{
The dynamical interpretations of $A$ are different for other powers of
$\frac{A}{r}$.
}
as the mass $M$.
Table~1 contains some astrophysically interesting masses that mimic
the values of the $A$ parameter.

\begin{table}[tbp]
\centering
\begin{tabular}{|p{35mm}|p{17mm}|p{20mm}|}
\hline
\begin{tabular}{c}
\\
{\bf Example} \\
\end{tabular}
&
\begin{tabular}{c}
\\
$M \; in \; M_{\odot}$
\end{tabular}
&
\begin{tabular}{c}
\\
$A \; in \; pc$
\end{tabular}
\\ \hline\hline
\begin{tabular}{c}
sun
\end{tabular}
&
\begin{tabular}{c}
1.
\end{tabular}
&
\begin{tabular}{c}
$10^{-13}$
\end{tabular}
\\ \hline
\begin{tabular}{c}
globular cluster
\end{tabular}
&
\begin{tabular}{c}
$10^4 - 10^6$
\end{tabular}
&
\begin{tabular}{c}
$ 10^{-9} - 10^{-7}$
\end{tabular}
\\ \hline
\begin{tabular}{c}
galactic nucleus
\end{tabular}
&
\begin{tabular}{c}
$10^7$
\end{tabular}
&
\begin{tabular}{c}
$ 10^{-6}$
\end{tabular}
\\ \hline
\begin{tabular}{c}
galaxy
\end{tabular}
&
\begin{tabular}{c}
$5. \; 10^{11}$
\end{tabular}
&
\begin{tabular}{c}
$ 2. \; 10^{-2}$
\end{tabular}
\\ \hline
\begin{tabular}{c}
galaxy--quasar system
\end{tabular}
&
\begin{tabular}{c}
$5. \; 10^{18}$
\end{tabular}
&
\begin{tabular}{c}
$ 2. \; 10^5$
\end{tabular}
\\ \hline\hline
\end{tabular}
\caption{\label{prop-table}
{\small Values of the $A$ parameter for which the
six-dimensionality of the world influences the dynamics of test particles
in a similar way as the existence (in the 4-dimensional world) of mass $M$
given for some examples motivated by astrophysics.The third example and
the last example are exceptional in the sense that
the $A$ parameter has been estimated by demand to explain the observed
redshift peculiarities of such systems. Hence the mass $M$ has
purely effective meaning here --- for details see
Sections~4~and~5.}}
\end{table}

In this case the gravitational potential $g_{tt}$ (see Eq.(\ref{row_26}) or
Eq.(\ref{row_31})) is attractive although there is no massive matter acting
as a source.

Let us recall that, in standard derivation of the Schwarzschild solution, the
free parameter in the metric tensor is identified with the total mass of
spherically symmetric configuration by the demand that at large distances the
metric tensor should reproduce the Newtonian potential. In our case, we cannot
identify the $A$ parameter directly with $M.$ The reason is that our solution
describes the case where ordinary matter is absent. The only contribution to
the energy-momentum tensor comes from the dilaton (massless) scalar field
$\varphi.$

It is well known (see \cite{Land-Lif}) that the frequency $\omega_{0}$ of
light, moving along the geodesic line in gravitational field which is static
or stationary, measured in the units of time $t$, is constant ($\omega_{0} =
constant$) along the geodesic. The frequency $\omega$ of light as a function
of the proper time~$\tau$ ($d \tau = \sqrt{g_{tt}} \: dt$ ) is equal to
\begin{eqnarray}  \label{row_43}
\omega = \omega_{0} \: \frac{dt}{d\tau} = \frac{\omega_{0}}{\sqrt{g_{tt}}} =
\omega_{0} \: \sqrt{g^{tt}} \; \; .
\end{eqnarray}

Let us assume that photon with frequency $\omega_{s}$ (measured in units of
the proper time $\tau$ ) is emitted
from the source which is located at a point $r = r_{s}$ where $g_{tt} =
g^{s}_{tt}$. Then the photon is moving along a geodesic line and it reaches the
observer at the point $r = r_{obs}$, where $g_{tt} = g^{obs}_{tt}$, with the
frequency $\omega_{obs} $
\begin{eqnarray}  \label{row_44}
\frac{\omega_{obs}}{\omega_{s}} = \sqrt{\frac{g_{obs}^{tt}}{g_{s}^{tt}}} =
\sqrt{\frac{g^{s}_{tt}}{g^{obs}_{tt}}} \; \; .
\end{eqnarray}
Using Eq.(\ref{row_31}) we can rewrite Eq.(\ref{row_44}) as
\begin{eqnarray}  \label{row_45}
\frac{\omega_{obs}}{\omega_{s}} = \frac{\sqrt{\frac{r_{s}}{r_{s} + A}}}{
\sqrt{\frac{r_{obs}}{r_{obs} + A}}} \;.
\end{eqnarray}
For simplicity, let us consider the limiting case when the observer is
situated at the infinity. Then we get
\begin{eqnarray}  \label{row_46}
\frac{\omega_{obs}}{\omega_{s}} = \sqrt{\frac{r^{w}_{s}}{r^{w}_{s} + 1}} \;
\;\; ,  {\rm where} \; \;\; r^{w}_{s} = \frac{r_{s}}{A} \; \; .
\end{eqnarray}
Therefore, we obtain that the nearer is the source to the center of the field
$\varphi (r) $ the more is the emitted photon redshifted at the point where
it reaches the observer (It should also be emphasized that this redshift does
depend on the relative radius $r^{w} = \frac{r}{A} $ rather than separately
on $r$ and $A$ (see also Appendix~A.2)).

\section{Six-dimensional Hamilton-Jacobi equation for a ``test particle''}
\label{motion}

In order to gain better understanding concerning possible manifestations of
six-dimensionality of the world
let us investigate the motion of a ``particle'' with a mass $m \neq 0 $ in the
central gravitational field described by the Eq.(\ref{row_26}).  Whether the
moving object can be treated as a ``test particle'' depends on the value of
the $A$ parameter in the metric tensor $g_{MN}$
(see Eq.(\ref{row_26})). Here $m $ is the mass of a ``test particle'' in the
six - dimensional spacetime.

The Hamilton-Jacobi equation
\begin{eqnarray}  \label{row_49}
g^{MN} \: \partial_{M} S \, \partial_{N} S - m^{2} c^{2} = 0
\end{eqnarray}
describing the motion of a test particle reads
{\small
\begin{eqnarray}  \label{row_50}
\!\!\!\!\!\!\!\!\!\!&& \frac{r + A}{r}\left( \frac{\partial S}{c \partial t} \right)^{2} -
\frac{r + A}{r} \: \left( \frac{\partial S}{\partial r} \right)^{2} -
\frac{1}{r^{2}} \: \left( \frac{\partial S}{\partial \Theta} \right)^{2} - \nonumber \\
&-& \frac{1}{r^{2} \: sin^{2} \Theta} \: \left( \frac{\partial S}{\partial \Phi} \right)^{2}
- \frac{1}{d^{2}} \: \frac{r}{r + A} \: \frac{1}{cos^{2} \vartheta} \:
\left( \frac{\partial S}{\partial \vartheta} \right)^{2} - \nonumber \\
&-& \frac{1}{d^{2}} \: \frac{r}{r + A} \: \left( \frac{\partial S}{\partial \varsigma}
\right)^{2} - m^{2} \: c^{2} = 0 \; ,
\end{eqnarray}
}
where $S $ denotes the action\footnote{
The action $S$ has the meaning of the complete integral of the Hamilton -
Jacobi equation and should not be confused with six-dimensional
field-theoretical action ${\cal S}$ invoked in
Section~2.}.

Without lack of generality we shall restrict ourselves to the motion in the
plane $\Theta = {\pi}/2$, thus
\begin{eqnarray}  \label{row_51}
\frac{\partial S}{\partial \Theta} = 0 \; \; .
\end{eqnarray}

The standard procedure of the separation of variables begins with the
following factorization of $S $
\begin{eqnarray}  \label{row_52}
S = - {\cal E}_{0} \: t + {\cal M}_{\Phi} \: \Phi + S_{r} (r) + {\cal M}
_{\varsigma} \: \varsigma + S_{\vartheta} (\vartheta)\;.
\end{eqnarray}
Then by separating Eq.(\ref{row_50}) into four-dimensional and internal
parts one arrives at the formula
{\small
\begin{eqnarray}  \label{separation}
\!\!\!\!\!&&\frac{r + A}{r} \: \left[ \frac{r + A}{r} \: (\frac{{\cal E}_{0}}{c})^{2} -
\frac{r + A}{r} \: (\frac{\partial S_{r}}{\partial r})^{2} - \right. \nonumber \\
&-& \left. \frac{1}{r^{2} \: sin^{2} \Theta} \: {\cal M}_{\Phi}^{2} - m^{2} \: c^{2} \right] =  \\
&=& \frac{1}{d^{2}} \: \frac{1}{cos^{2} \vartheta} \: ( \frac{\partial
S_{\vartheta}}{\partial \vartheta})^{2} + \frac{1}{d^{2}} \: {\cal M}
_{\varsigma}^{2} =: k_{\vartheta \varsigma}^{2} = constant \: . \nonumber
\end{eqnarray}
}
The last equation in (\ref{separation}) is easy to integrate for
$S_{\vartheta}$
\begin{eqnarray}  \label{row_53}
S_{\vartheta} = {\pm} \: ( d^{2} \: k_{\vartheta \varsigma}^{2} - {\cal M}
_{\varsigma}^{2} )^{\frac{1}{2}} \: sin \vartheta = {\pm} \: k_{\vartheta}
\: sin \vartheta. \;
\end{eqnarray}
The separation constants ${\cal E}_{0}$, ${\cal M}_{\Phi}$, ${\cal M}
_{\varsigma}$ and $k_{\vartheta} $ have the meaning of the total energy, the
effectual angular momentum, the internal angular momentum and
internal momentum respectively. One may also distinguish the internal total
momentum $k_{\vartheta \varsigma}$
\begin{eqnarray}  \label{row_54}
k_{\vartheta \varsigma}^{2} = \frac{{\cal M}_{\varsigma}^{2}}{d^{2}} +
k_{\vartheta}^{2} \; \; .
\end{eqnarray}
Analogously the quantity
\begin{eqnarray}  \label{row_55}
m_{4}^{2} = m^{2} + \frac{k_{\vartheta \varsigma}^{2}}{c^2}
\end{eqnarray}
can be interpreted as the four - dimensional squared mass of a test particle
in the flat
Min\-kow\-skian limit at infinity. In the case of vanishing internal
momentum $k_{\vartheta \varsigma} = 0$ the four and six-dimensional masses
are equal $m_{4} = m$. If the six-dimensional mass is zero $m = 0$ then the
four - dimensional mass at spatial infinity would be solely of internal
origin: $m_{4} = \frac{|k_{\vartheta \varsigma}|}{c} $.

Let us now consider the radial part $S_{r} (r) $ which can be easily read
off from the Eq.(\ref{separation})
\begin{eqnarray}  \label{row_56}
\frac{d S_{r}}{d r} &=&  \left[ \frac{{\cal E}_{0}^2}{c^2} - \left( m^2 \, c^2 +
\frac{{\cal M}_{\Phi}^2}{r^2} \right) \,
\frac{r}{r + A} - \right. \nonumber \\
&-& \left. k_{\vartheta \varsigma}^2 \, \left( \frac{r}{r + A} \right)^2 \right]^{\frac{1}{2}}
\end{eqnarray}
or after formal integration
\begin{eqnarray}  \label{row_57}
S_{r} (r) &=& \int dr\, \left[ \frac{{\cal E}_{0}^2}{c^2} -
\left( m^2 \, c^2 + \frac{{\cal M}_{\Phi}^2}{r^2} \right) \,
\frac{r}{r+ A} - \right. \nonumber \\
&-& \left. k_{\vartheta \varsigma}^2 \, \left( \frac{r}{r + A} \right)^2 \right]^{\frac{1}{2}} \; \; .
\end{eqnarray}
The trajectory of a test particle is implicitly determined by the following equations
\begin{eqnarray}  \label{row_58}
\frac{\partial S}{\partial {\cal E}_{0}} = \alpha_{1} = - t + \frac{\partial
S_{r} (r)}{\partial {\cal E}_{0}}
\end{eqnarray}
\begin{eqnarray}  \label{row_59}
\frac{\partial S}{\partial {\cal M}_{\Phi}} = \alpha_{2} = \Phi + \frac{
\partial S_{r} (r)}{\partial {\cal M}_{\Phi}}
\end{eqnarray}
where $\alpha_{1}$ and $\alpha_{2}$ are constants and without lack of
generality the initial conditions can be chosen so that $\alpha_{1} =
\alpha_{2} = 0 .$ In other words integration of Eq.(\ref{row_59}) gives
the trajectory $r = r (\Phi)$ of a test particle and Eq.(\ref{row_58})
provides the temporal dependence of radial coordinate $r = r (t).$ These two
relations determine the trajectory of a test particle $r = r (t)$ and $\Phi
= \Phi (t)$ \cite{Rubinowicz-Krolikowski-1980}.

Implementing the above outlined procedure we obtain
(from Eq.(\ref{row_58}) and  Eq.(\ref{row_57}))
\begin{eqnarray}  \label{row_60}
t &=& \frac{{\cal E}_{0}}{c^2}\,\int dr \left[ {\frac{{{{{\cal E}_{0}}^2}}}{{
c^2}}} - {\left( {m^2}\,{c^2} + {\frac{{{{{\cal M}_{\Phi}}^2}}}{{r^2}}}
\right) \, \frac{r}{r + A}} - \right. \nonumber \\
&-& \left. {k_{\vartheta \varsigma}^2} \, {{{\ \left(
\frac{r}{r + A} \right) }^2}} \right]^{- \frac{1}{2}}
\end{eqnarray}
and consequently the radial velocity
\begin{eqnarray}  \label{row_61}
\frac{d r}{d t} &=&  \frac{c^2}{{\cal E}_{0}} \,
\left[{\frac{ {{\ {{\cal E}_{0}}^2}}}{{c^2} }} - \left( {m^2}\,{c^2} + {\frac{{{{{\cal M}_{\Phi}}^2}}}{{r^2}
}} \right) \, \frac{r}{r + A} - \right. \nonumber \\
&-& \left. {k_{\vartheta \varsigma}^2} \, {{{\ \left(\frac{r}{r + A} \right) }^2}} \right]^{\frac{1}{2}} \; \; .
\end{eqnarray}
It is easy to notice that the quantity
\begin{eqnarray}  \label{row_62}
m_{4}(r) = \sqrt{ m^2 \, \left( \frac{r}{r + A} \right) + \frac{k_{\vartheta
\varsigma}^2}{c^2} \, \left( \frac{r}{r + A} \right)^2 }
\end{eqnarray}
can be interpreted as the mass in the four - dimensional curved
spacetime.
The $m_{4}$ of Eq.(\ref{row_55}) is recovered as the limit of $
m_{4}(r)$ for $r \rightarrow \infty$. Similarly to our previous discussion,
if $k_{\vartheta \varsigma} = 0$ then $m_{4}(r) = m \, \sqrt{\frac{r}{r + A}}
$ and if $m = 0$ then the mass $m_{4}(r)$ would have the internal origin $
m_{4}(r) = \frac{|k_{\vartheta \varsigma}|}{c} \, \left( \frac{r}{r + A}
\right) $. It should also be emphasized that $m_{4}(r)$ does depend on the
ratio $r^{w} = \frac{r}{A} $ rather than separately on $r$ and $A$,
reflecting the scale-invariance of the model (see also Appendix~A.2).

In a similar manner using the Eq.(\ref{row_59}) and Eq.(\ref{row_57}) we
obtain
\begin{eqnarray}  \label{row_63}
\Phi &=& \int \left[{\frac{{{{{\cal E}_{0}}^2}}}{{c^2}}} - {\left( {m^2}\,{c^2}
+ {\frac{{{{{\cal M}_{\Phi}}^2}}}{{r^2}}} \right) \frac{r}{r + A}} - \right. \nonumber \\
&-& \left. \left( {k_{\vartheta \varsigma}^2} \right) {{{\ \left( \frac{r}{r + A}
\right) }^2}} \right]^{- \frac{1}{2}} \,
\frac{{\cal M}_{\Phi} \, dr}{r\,(r + A)}
\end{eqnarray}
and consequently using Eqs.(\ref{row_63}) and (\ref{row_61}) the angular
velocity of the particle reads
\begin{eqnarray}  \label{row_64}
{\Omega}_{t} = {\frac{d \Phi}{d t}} = {\frac{d \Phi}{d r}} \,\frac{d r}{d t}
= {\frac{{{c^2}\,{\cal M}_{\Phi}}}{{{\cal E}_{0} \, r \,\left( A + r \right)
}}} \; \; .
\end{eqnarray}
Now the proper angular velocity of the particle is equal to
\begin{eqnarray}  \label{row_65}
{\Omega}_{\tau} = {\frac{d \Phi}{d \tau}} = {\Omega}_{t} \, \sqrt{\frac{r + A
}{r}} \; ,
\end{eqnarray}
where $\tau $ is the proper time and $d \tau = \sqrt{g_{tt}} \,dt $ (see
Eq.(\ref{row_26})).

The transversal velocity of the particle (i.e. the component perpendicular
to the radial direction) is equal to (see Eq.(\ref{row_32}))
\begin{eqnarray}  \label{row_66}
v_{t} = {\Omega}_{t} \, r_{l}
\end{eqnarray}
and analogously the transversal component of the proper velocity (written in
the units of proper time $\tau$ ) is equal to
\begin{eqnarray}  \label{row_67}
v_{\tau} = {\Omega}_{\tau} \, r_{l} = v_{t} \, \sqrt{\frac{r + A}{r}}
\end{eqnarray}
where ${\Omega}_{t} $, ${\Omega}_{\tau} $ and $r_{l} $ are given by
Eqs.(\ref{row_64}),(\ref{row_65})~and~(\ref{row_32}) respectively.

Let us rewrite Eq.(\ref{row_61}) in the following form
\begin{eqnarray}  \label{row_68}
\frac{dr}{dt} = \frac{c}{{\cal E}_{0}} \: \sqrt{{\cal E}_{0}^{2} -
{\cal U}_{eff}^{2} (r) }
\end{eqnarray}
where
\begin{eqnarray}  \label{row_69}
{\cal U}_{eff} (r) &=& \left[ \left( m^{2} \, c^{4} +
\left( \frac{{\cal M}_{\Phi}}{r} \right)^{2} \, c^{2} \right) \,
\frac{r}{r + A} + \right. \nonumber \\
&+& \left. \left( k_{\vartheta \varsigma}^{2} \,c^{2} \right)
\left( \frac{r}{r + A} \right)^{2} \, \right]^{\frac{1}{2}} \; \; .
\end{eqnarray}
The function ${\cal U}_{eff} (r) $ plays the role of the effective potential
energy in the sense that the relation between ${\cal E}_{0}$ and
${\cal U}_{eff} (r)$ determines the allowed regions of the motion of the
particle.

The proper radial velocity of the particle is equal to
\begin{eqnarray}  \label{row_70}
v_{r} = \frac{d r_{l}}{d \tau} = \frac{\sqrt{- g_{rr}} \, d r}{\sqrt{g_{tt}}
d t} = \frac{d r}{d t} \; ,
\end{eqnarray}
where in the last equality we used the relation $g_{tt} = - g_{rr} $ (see
Eq.(\ref{row_31})) and $\tau $ is the proper time. So the radial velocity
$dr/dt$ and the proper radial velocity $v_{r} = dr_{l}/d \tau $ are equal and
one should stress that this property is fundamentally different from the
analogical relation for the black hole.

\subsection{Stable circular orbits.}

Let us now consider the motion of a particle with given values of
${\cal M}_{\Phi} \neq 0 $ and $k_{\vartheta \varsigma}$.
Figure~1 illustrates the function ${\cal U}_{eff}(r) $ for different values of
${\cal M}_{\Phi} $ and $k_{\vartheta \varsigma} = 0.$

Similarly Figure~2 shows the function ${\cal U}_{eff} (r) $ for
different values of $k_{\vartheta \varsigma} $ with fixed value
of ${\cal M}_{\Phi}.$

Let us consider a stable circular orbit with given values of
${\cal E}_{0}$, ${\cal M}_{\Phi}$ and $k_{\vartheta \varsigma} $.
The radius of this orbit can be calculated from the following
equations (see Eqs.(\ref{row_68}),  (\ref{row_69}))
\begin{eqnarray}  \label{row_74}
\frac{dr}{dt} = 0 \; ,
\end{eqnarray}
\begin{eqnarray}  \label{row_75}
\frac{d^{2} r}{dt^{2}} = \frac{dr}{dt} \frac{d}{dr}(\frac{dr}{dt}) = 0 \; .
\end{eqnarray}
From Eqs.(\ref{row_75}),(\ref{row_68}),(\ref{row_69}) we obtain the implicit
relation between the radius $r$ of the stable circular orbit, the angular
momentum of the particle ${\cal M}_{\Phi}$ and the internal ``total momentum''
$k_{\vartheta \varsigma}$
\begin{eqnarray}  \label{row_76}
({\cal M}_{\Phi})^{2} = \left( m^2 \,c^2 + 2 \, k_{\vartheta
\varsigma}^{2} ( \frac{r}{r + A} ) \right) \left( \frac{A}{2 \, r + A}
\right) r^{2}
\end{eqnarray}
so
\begin{eqnarray}  \label{row_77}
{\cal M}_{\Phi} = \pm A^{1/2} \, r \,\sqrt{\frac{(r + A) \, m^2 \, c^2
+ 2 \, k_{\vartheta \varsigma}^{2} \, r }{A^2 + 3 \,A \, r + 2 \,r^2}} \; \; .
\end{eqnarray}
It is easy to verify that ${\cal M}_{\Phi} / r \rightarrow 0$ as $r$
tends to infinity. Using Eqs.(\ref{row_74}),(\ref{row_68}),  (\ref{row_69}) we
calculate the total energy ${\cal E}_{0}$ of the particle moving along
this stable circular orbit of radius $r$
\begin{eqnarray}  \label{row_78}
({\cal E}_{0})^{2} & = & \left[ m^2 \,c^4 + \left( m^2 \,c^4 + \right. \right. \nonumber \\
&+& \left. \left. 2 \, k_{\vartheta \varsigma}^{2} \,c^2 ( \frac{r}{r + A} ) \right) \frac{A}{2 \, r + A} \right]
\left( \frac{r}{r + A} \right) +  \nonumber \\
& + & k_{\vartheta \varsigma}^{2} c^{2} \left( \frac{r}{r + A} \right)^2
\end{eqnarray}
or in other words
{\small
\begin{eqnarray}  \label{row_79}
\!\!\!\!\!{\cal E}_{0} = c^2 \,\sqrt{\frac{ r\, (2 \, (r + A)^{2} \,m^2 \,c^2 + 3
\, A \, r\,k_{\vartheta \varsigma}^{2} + 2 \, r^2 \, k_{\vartheta
\varsigma}^2 )} {c^2 \, (r + A)^2 \, (2 \,r + A) } } \:  .
\end{eqnarray}
}

There are some physically obvious conditions which in turn restrict
admissible values of radii of stable orbits for given internal momentum
$k_{\vartheta \varsigma},$ namely
\begin{eqnarray}  \label{row_80}
\left( {\cal M}_{\Phi} \right)^{2} > 0 \; \; , \; {\rm so} \; \; \;
k_{\vartheta \varsigma}^2 > - \frac{m^2 \,c^2}{2} \, ( \frac{r + A}{r} )
\end{eqnarray}
\begin{eqnarray}  \label{row_81}
\left( {\cal E}_{0} \right)^{2} \geq 0 \; \; , \; {\rm so} \; \;
k_{\vartheta \varsigma}^2 \geq - m^2 \,c^2 \,\frac{2 \,(r + A)}{3 \,A + 2 \,r}
 \, ( \frac{r + A}{r} ) \: .
\end{eqnarray}
It is not difficult to verify that condition (\ref{row_80}) is stronger than
(\ref{row_81}).

Now let us calculate the proper angular velocity, and the proper transversal
velocity of the particle moving along the stable circular orbit given by
Eqs.(\ref{row_74}),(\ref{row_75}). Using Eqs.(\ref{row_64})-(\ref{row_67})
with ${\cal M}_{\Phi}$ and ${\cal E}_{0}$ given by Eqs.(\ref{row_77}),
(\ref{row_79}) respectively, we obtain the proper angular velocity
\begin{eqnarray}  \label{row_82}
\Omega_{\tau}^{st} = {\frac{{{c^2}\,{\cal M}_{\Phi}}}{{{\cal E}_{0} \, r \,
\left( A + r \right) }}}\, \sqrt{\frac{r + A}{r}}
\end{eqnarray}
and the proper tangent velocity
\begin{eqnarray}  \label{row_83}
v_{\tau}^{st} = {\frac{{{c^2}\,{\cal M}_{\Phi}}}{{{\cal E}_{0} \, r \,
\left( A + r \right) }}}\,r_{l}\, \sqrt{\frac{r + A}{r}} \; \;
\end{eqnarray}
on the stable (index $st$) circular orbit. \\
One can notice from Eq.(\ref{row_83}) and Eqs.(\ref{row_77}),  (\ref{row_79})
that if $k_{\vartheta \varsigma}^{2} \propto Constant \cdot m^2 $ then
$v_{\tau}^{st}$ does not depend explicitly on mass $m$ of the particle.
However $v_{\tau}^{st}$ is still parameterized by the value of proportionality
constant which is clearly non-classical effect.
It is also not difficult to see that $v_{\tau}^{st}$ again depends on the
ratio $r^{w} = \frac{r}{A} $ rather then on $r$ and $A$ independently.

\subsubsection{The case with $m \neq 0$.}

From Eq.(\ref{row_80}) (see Figure 3) we may notice
that if $k_{\vartheta \varsigma}^{2} > - \frac{1}{2} \, m^2 \, c^2 $
then all values of the radius $r $ (or $r_{l} $ cf. Eq.(\ref{row_32})) for
stable orbits are allowed.

If however $k_{\vartheta \varsigma,min}^{2} \leq k_{\vartheta \varsigma}^{2} <
- \frac{1}{2} \, m^2 \, c^2 $
(the lower limit $k_{\vartheta \varsigma,min}^{2}$ is till now undefined)
then according to Eq.(\ref{row_80}) the stable orbits may exist only up to a
radius
\begin{equation}  \label{rmax1}
r_{0} \equiv - A/(1 + \frac{2\,k_{\vartheta \varsigma}^2}{m^2 \, c^2}) \, .
\end{equation}
In other words Eq.(\ref{row_80}) implies that one cannot find any stable
orbit with $r \geq r_{0}$ for fixed $k_{\vartheta \varsigma}$ and the angular
momentum ${\cal M}_{\Phi}$ chosen according to Eq.(\ref{row_77}).
From the demand that the value of the effective potential at infinity can be
lowered only up to the value of the local minimum of the effective potential
at radius $r$, it follows that the lower limit
$k_{\vartheta \varsigma,min}^{2}$ (which appears to depend on the radius $r$
($r > 0$)) is equal to
\begin{equation}  \label{lowerlimit}
k_{\vartheta \varsigma,min}^{2}(r) \equiv - m^2 c^2 \frac{(1 + \frac{r}{A})^2}
{1 + 4\frac{r}{A} + 2(\frac{r}{A})^2} > - m^2 c^2 \, .
\end{equation}
Now from Eq.(\ref{lowerlimit}) we obtain another limit for allowed stable
orbits (in the case of $k_{\vartheta \varsigma,min}^{2} \leq
k_{\vartheta \varsigma}^{2} < - \frac{1}{2} \, m^2 \, c^2 $)
{\small
\begin{equation}  \label{rmax2}
\!\!\!\!\! 0 < r_{max} \equiv r_{0} - A \frac{ 2 \, k_{\vartheta \varsigma}^2 +
\sqrt{k_{\vartheta \varsigma}^2 (2 \,k_{\vartheta \varsigma}^2 + m^{2} c^{2})}}
{2 \, k_{\vartheta \varsigma}^2 + m^{2} c^{2}} < r_{0} \, ,
\end{equation}}
so, the maximal allowed radius of the stable orbit is  equal to $r_{max}$
which is smaller than $r_{0}$.

Figure 3 displays the rotation curves calculated
according to Eq.(\ref{row_83}) (in units of the velocity of light c) for different
values of $k_{\vartheta \varsigma}$.
It shows that the whole effect, mostly pronounced in regions
close to the center of spherically symmetric configurations of the field
$\varphi$, is extended from the center to the infinity.
(The value of contribution to total rotational velocity
coming from the scalar field $\varphi$ for the Sun at a distance of
$7.5 \, kpc $ from the center of the scalar field in our Galaxy
for $k_{\vartheta \varsigma} = 0$ is equal to $ 1.73 \, km/s$.)

\subsubsection{The case with $m = 0 $ and $k_{\vartheta \varsigma} \neq 0$.}

In the case $m = 0 $ and $k_{\vartheta \varsigma} \neq 0$ we can see from
the Eqs.(\ref{row_80})~and~(\ref{row_81}) that $k_{\vartheta \varsigma}^{2}
> 0$ and according to our previous discussion all values of $r$ are allowed.

\subsection{Radial trajectories.}

In this case we shall investigate free motion of a test particle (with given
internal ``total momentum'' $k_{\vartheta \varsigma} $) along the geodesic $
\Phi = constant $ (${\cal M}_{\Phi} = 0$)
crossing the center of the gravitational field $g_{MN} $. From the
Eqs.(\ref{row_68}),(\ref{row_69})~and~(\ref{row_70}) we obtain
\begin{eqnarray}  \label{row_71}
\!\!\!\!\!\!\!\!\!\!
&&v_{r} = \frac{dr}{dt} = \nonumber \\
\!\!\!\!\!\!\!\!\!\!
&=& \frac{c}{{\cal E}_{0}} \, \sqrt{{\cal E}_{0}^{2} -
\left( {m^2}\,{c^4} \right) \, \frac{r}{r + A} - k_{\vartheta \varsigma}^2
\, c^2 \, \left( \frac{r}{r + A} \right)^2 } \: .
\end{eqnarray}
From this equation we may see that for the particle which is initially
($t=t_{o}$) at rest ($v_{r}=v_{r_{o}}=0$) at the point $r = r_{o} \neq 0$ the
total energy is equal to (compare Eq.(\ref{row_62}))
\begin{eqnarray}  \label{row_72}
{\cal E}_{0} &=& \sqrt{ m^2 \,c^4 \, \left( \frac{r_{o}}{r_{o} + A} \right) +
k_{\vartheta \varsigma}^2 \,c^2 \,(\frac{r_{o}}{r_{o} + A})^2 \, } = \nonumber \\
&=& m_{4}(r_{o}) \, c^{2}.
\end{eqnarray}
Hence the particle is oscillating and crosses the center with velocity $
v_{r}(r=0) = c $ (at the center the particle becomes massless $m_{4}(r=0)=0$).
The acceleration of the particle is equal to
\begin{eqnarray}  \label{row_73}
a_{r} &=& \frac{d v_{r}}{d \tau} = \nonumber \\
&=& - A \,c^2 \,\frac{ m^2 \,c^4 +
2 \,k_{\vartheta \varsigma}^2 \, c^2 \,\left( \frac{r}{r + A} \right) }
{2 \, {\cal E}_{0}^{2} \, (r + A)^2} \sqrt{\frac{r + A}{r}} \; .
\end{eqnarray}
Because the potential energy at infinity cannot be lower than at the center
at $r=0$, then simple comparison of these limits based on formula
(\ref{row_69}) reveals that $k_{\vartheta \varsigma}^2 \geq - m^{2} \,c^{2}$,
hence $m_{4}^{2}(r) \geq 0$ everywhere\footnote{
Because in this case the effective potential is globally the lower one,
it follows from this condition that the proportionality constant in equation
$k_{\vartheta \varsigma}^{2} \propto Constant \cdot m^2 $ should be equal to
$(- c^{2})$. Hence we obtain $k_{\vartheta \varsigma}^{2} = - c^{2}
\cdot m^2 $ as the lower value for all test particles with $m^{2} \neq 0$,
which we \cite{Dziekuje Ci Panie Jezu Chryste} believe to be the fundamental
law of nature - which we call - the law of the lower potential.
}
(cf. Eq.(\ref{row_62})).

{\it Note:
A remark is in order at this point. Namely it may appear obvious that
$k_{\vartheta \varsigma}^2 \geq - m^{2} \,c^{2}$ (so we admit imaginary
values of $k_{\vartheta \varsigma}$). However it is our conviction that one
should not uncritically transfer four-dimensional intuitions (such like
$k_{\vartheta \varsigma}^2 \geq 0$ or $m^2 \geq 0$, although some of them may
turn out to be true) but rather build on safe grounds of known properties of
four-dimensional sector i.e. $m_4^2 \geq 0$ in this case.}

\subsubsection{The case with $m \neq 0$.}

From Eq.(\ref{row_73}) we see that the acceleration tends to minus infinity
at the center and when $k_{\vartheta \varsigma}^{2} > - \frac{1}{2} \,
m^2 \, c^2 $ it monotonously increases to the zero value when $r$ is going
to infinity. So, in this case, the particle is attracted to the center for
all values of $r$ (see Figure 4).

In the case of $- m^{2} \, c^{2} \leq k_{\vartheta \varsigma}^{2}
< - \frac{1}{2} \, m^2 \, c^2 $ there exists the finite value of
$r = r_{a_{r} = 0}$ for which $a_{r} = 0$. For $r \leq r_{o} <
r_{a_{r} = 0}$ the particle is attracted to the center with
$a_{r} \rightarrow - \infty $ for $r \rightarrow 0 $ and for $r
\geq r_{o} > r_{a_{r} = 0}$ the particle is repelled from the
center and the acceleration $a_{r} \rightarrow 0$ when $r
\rightarrow \infty$ (see Figure 4). On Figure 5 the radial
acceleration $a_{r}$ of the particle which is very closely to the
center of the field $\varphi$ is presented. Because the solution
is horizon free the achieved values of acceleration of the
particle would be the cause of a visible point like radiation (see
Section~5).

\subsubsection{ The case with $m = 0 $ and $k_{\vartheta \varsigma} \neq 0$.}

In this case the requirement of ${\cal E}_{0}^{2} > 0$ implies $k_{\vartheta
\varsigma}^{2} > 0$ (see Eq.(\ref{row_72})). From Eqs.(\ref{row_71}),
(\ref{row_72})~and~(\ref{row_73}) we conclude
that the particle is attracted to the center ($a_{r} < 0$) for all values of
$r$.

\subsection{ The case with $m = 0 $ and $k_{\vartheta \varsigma} = 0$.}

Now we have ${\cal E}_{0} = constant \neq 0$ and one can read from
Eqs.(\ref{row_71})~and~(\ref{row_73}) that $v_{r} = c$ and $a_{r} = 0$ for all
values of $r$.
So the particle which has both the six-dimensional mass $m$ and the
internal ``total momentum'' $k_{\vartheta \varsigma}$ equal to zero (which is
a reasonable representation of a photon for example) does not feel
(except of changing the frequency, see Eq.(\ref{row_43})) the curvature of
the spacetime when moving along the geodesic line crossing the center.
On the other hand for ${\cal M}_{\Phi} \neq 0 $, $m = 0 $ and
$k_{\vartheta \varsigma} = 0$ by using the Eq.(\ref{row_63}) and introducing
formally the parameter\footnote{
We should use for $m = 0$ and $k_{\vartheta \varsigma} = 0$ the eikonal
equation instead of the Hamilton-Jacobi equation. However formal (technical)
substitution of $r_{m} = \frac{{\cal M}_{\Phi} \,c}{{\cal E}_{0}}$ gives the
same analytical result.} $r_{m} = \frac{{\cal M}_{\Phi} \,c}{{\cal E}_{0}}$
we obtain the trajectory of the particle
\begin{eqnarray}  \label{Phi-trajectory}
\Phi = \int \left[\frac{1}{r_{m}^{2}} - \frac{1}{r^{2}} \, \frac{r}{r + A}
\right]^{ - \frac{1}{2}} \frac{dr}{r\,(r + A)} \, \, .
\end{eqnarray}
When $A \rightarrow 0$ then the trajectory calculated according to above
equation is the straight line $r = r_{m}/(cos \Phi) $ passing by the center
at a distance of $r_{m}$ (impact parameter). On the other hand light
traveling in our spacetime (with $A\neq 0$) is deflected even in absence of
baryonic matter.

\subsection{Redshift of the radiation from stable circular orbits.}

Let us suppose for simplicity that distant observer is located far away from
the center of the system at a distance far bigger than the size of the
system, so its peculiar motion with respect to the center of the system is
negligibly small. Let us also assume that the dynamical time scale $t_{dyn}$
is greater than the characteristic timescale $t_{obs}$ over which the
observations are performed i.e. $t_{dyn} > t_{obs}.$ In such a case the
observed motion of the ``particle'' is seen only as instantaneous redshift or
blueshift. If the motion of a ``particle'' takes place along the stable
circular orbit the Doppler shift is equal to (see Figure 6)
\begin{eqnarray}  \label{row_z1}
z_{D} = \sqrt{\frac{c + v_{ \tau}^{st} sin \Phi ( \tau ) } {c -
v_{\tau}^{st} sin \Phi ( \tau )} } - 1 \; ,
\end{eqnarray}
where $\Phi ( \tau ) = \int_{0}^{\tau} \, \Omega_{ \tau}^{st} d \tau $ (see
Eq.(\ref{row_65}) and Eq.(\ref{row_82})) and the angle $\Phi (\tau)$ is
counted from the direction to observer i.e. $\Phi(\tau=0) = 0$ and
$v_{\tau}^{st}$ is given by Eq.(\ref{row_83}). Now the gravitational
redshift according to Eq.(\ref{row_46}) is equal to
(see Figure 5)
\begin{eqnarray}  \label{row_z2}
z_{g} = \frac{\lambda_{obs}}{\lambda_{s}} - 1 = \frac{\omega_{s}}{
\omega_{obs}} - 1 = \sqrt{\frac{r_{s} + A}{r_{s}}} - 1 \; \; .
\end{eqnarray}

It is not difficult to see that the combined effect of these
redshifts is the following (see Figure 7)
\begin{eqnarray}  \label{row_z3}
z = \left( z_{g} + 1 \right) \, \left( z_{D} + 1 \right) - 1 \, \, .
\end{eqnarray}

\section{Conclusions and perspectives}
\label{noncosm-concl}

In the present paper we have recognized certain non-perturbative
six-dimensional "spherically" symmetric solutions of the Einstein equations.
They are asymptotically flat but fundamentally different from the
Schwarz\-schild solutions in four-dimensional spacetime. The motion of test
particles has been analyzed.
The solutions presented in Section~3 are parameterized by the
parameter $A$ which has similar dynamical consequences (especially for
$r >> A$) as the mass {\small $M = \displaystyle{\frac{A c^2}{2 G}}$} --- its
existence would be perceived by an observer in the same way as invisible
mass.

We may imagine that our six-dimensional world could be compactified in
an non-homogeneous manner.
In this picture the scalar "basic" field $\varphi$ would form a kind of
ground field\footnote{\label{f-cyt-11}
It is interesting to notice that recently there has been
a strong tendency to draw analogies between cosmology and condensed matter
physics \cite{Hu-1996,Smolin-1995}.}.
The phenomenon of flat rotation curves and the connection of presented
model to the so called dark matter problem will be a subject of separate
paper.

However looking from the above mentioned perspective, we would like to invoke
here two classes of phenomena.
Some time ago Tanaka (1995) reported the detection of the
relativistic effects in an X-ray emission line (the $K\alpha$ line) from
ionized iron in the galaxy MCG-6-30-15.
The line is extremely broad, corresponding to a velocity of
$\sim 10^{5} \, km/s \approx 0.3\;c$, and asymmetric, with
most of the line flux being redshifted. This observation is not isolated
since, in several objects, broad redshifted lines have been detected but no
strong blue-shifted lines have been seen \cite{Tanaka-1995}.
This is an argument against any asymmetrical-outflow hypothesis in which
the flow is
directed away from us because some objects should then have the flow
directed towards us. On the other hand these observations have just been
properly explained (Section~4) for the motion along circular
orbits with the internal ``total momentum'' square $k_{\vartheta \varsigma}^2
< 0$ because in the model even the whole Doppler effect connected with the
blue wing moving towards us could be hidden bellow the gravitational redshift
effect (see Figures 6,7).

There are also other phenomena which call for explanation and are hard to
understand from the point of view of the standard lore. One class of such
problems is associated with the nature of the redshift of galaxies. There is
a number of evidence that the redshifts are at least partially intrinsic
properties of the galaxies, apparently quantized and time variable
\cite{Arp-1986,Tifft-1988,Tifft-93}.

It has been known for a long time that there exist associations of quasars
and galaxies where the components have widely discrepant redshifts
\cite{Arp-Giraud-Sulentic-Vigier-1983}.
We can think of a simple explanation of such apparently strange
phenomena in terms of presented model which has nothing to do with standard
cosmological interpretation of the redshift as a manifestation of the Hubble
expansion. Let us imagine that a quasar and galaxy system is captured by the
local configuration of the scalar field $\varphi$ -- such like described in
our model. It is natural to suppose that both the quasar and the galaxy
are moving along the stable orbits and we may thus apply the results obtained
in the Section~4.
If the things are arranged so that the quasar is closer to the center of the
$\varphi$ ground field configuration, it would have a (much) greater redshift
than the galaxy located peripherally. As it has been illustrated on
Figs.6,7 the whole idea works if the quasar is closer to the center than a
fraction of $A.$
At such distance the combined gravitational Doppler redshift given by
Eq.(\ref{row_z3}) is a steep function of $r$, which makes the idea working
if the quasar--galaxy constitute a close binary system.
Such a possibility has an attractive feature that bridges connecting galaxies
and quasars may be explained as evidence of an infall of matter from the
galaxy to the center of the field $\varphi$.
Let us recall that the field $\varphi$ strongly
accelerates the infalling particles (see Eq.(\ref{row_73})), and having in
mind our demand that the quasar is located closer to the center it may shed
some light on the nature of quasar emission. The idea briefly outlined above
deserves further deeper considerations. It would be also of interest to
examine interactions of these dilatonic centers and their distribution in
the universe.

The above mentioned observational facts \cite{Burbidge} cannot find any
reasonable explanation
within the standard interpretation which claims that the Hubble expansion is
responsible for the redshifts of galaxies. Hence, they require the search for
models entirely different from evolutionary ones. The true view of the
universe is not so faraway from us.

\section{Appendix~A.1: The solution for $A < 0$}
\label{noncosm-app}

If the parameter $A < 0$ then Eqs.(\ref{row_20})-(\ref{row_22}) are valid
only when $r > |A|$. The metric tensor becomes singular for $r = |A|.$
However its determinant $g$ (see Eq.(\ref{row_27})) remains well defined.
As in Section~3 we write the temporal and radial
components of the metric $g_{MN}$ and the internal ``radius'' $\varrho (r)$
(see Eqs.(\ref{row_21})~and~(\ref{row_26}))
\begin{eqnarray}  \label{row_36}
\left\{
\begin{array}{lll}
g_{tt} = \frac{r}{r - |A|} \; &  &  \\
g_{rr} = - \frac{r}{r - |A|} \; &  &  \\
\varrho (r) = d_{out} \sqrt{\frac{r - |A|}{r}} \; \; . &  &
\end{array}
\right.
\end{eqnarray}
The formulae for the scalar curvature ${\cal R}$ (see Eq.(\ref{row_24})) and
scalar field $\varphi$ Eq.(\ref{row_22}) are
\begin{eqnarray}  \label{row_38}
{\cal R} = \frac{A^{2}}{2 r^{3} (r - |A|)} \;
\end{eqnarray}
\begin{eqnarray}  \label{row_39}
\varphi (r) = \pm \, \sqrt{\frac{1}{2 \kappa_{6}}} \: \ln(\frac{r}{r - |A|})
\; ,
\end{eqnarray}
where $d_{out} $ is the constant.

Now the physical radial distance $r_{l-A}$ from the the radius $|A| $ to the
radius $r > |A| $ is equal to
{\small
\begin{eqnarray}  \label{row_37}
\!\!\!\!\!\!\!\!\!\!\!\! &r_{l-A}& = \int_{|A|}^{r} dr \, \sqrt{- g_{rr}} = \sqrt{\frac{r}{ r - |A|
}} \, (r - |A|) +   \\
\!\!\!\!\!\!\!\!\!\!\!\!\!\!\!\!\!\!& + & \!\!\!\! \frac{1}{2} \, |A| \, \ln \left( \frac{ - |A| + 2 r + 2 \, ( r - |A| )
\, \sqrt{\frac{r}{r - |A|}}}{|A|} \right) > r - |A|. \nonumber
\end{eqnarray}
}

Like in Section~2 we have $g_{tt} \rightarrow 1$
for $r \rightarrow \infty$ (see Eq.(\ref{row_31})).
Hence comparing the gravitational potential $g_{tt}
= \frac{r}{r - |A|} \approx 1 + \frac{|A|}{r}$ for $r\gg |A|$ with the
gravitational potential $g_{tt} = 1 - \frac{G}{c^{2}} \frac{M}{r} $ for a
field induced by the mass $M$ we notice, that gravitational potential $g_{tt}
$ (see Eq.(\ref{row_36})) is the repulsive one. This is the reason why
this case has been omitted in the main text as till now unobserved but because
it formally provides a solution to the model we reproduce some of its
properties here.

Now the real radial distance $r_{l}$ from the center $r = 0$ to the point with
$r > |A| $ is equal to (see Eq.(\ref{row_37}))
\begin{eqnarray}  \label{row_42}
r_{l} = |A| + r_{l-A} > r \; \; .
\end{eqnarray}
Using Eq.(\ref{row_36}) we can rewrite Eq.(\ref{row_44}) as follows
\begin{eqnarray}  \label{row_47}
\frac{\omega_{ob}}{\omega_{s}} = \frac{\sqrt{\frac{r_{s}}{r_{s} - |A|}}}{
\sqrt{\frac{r_{obs}}{r_{obs} - |A|}}} \; \; .
\end{eqnarray}
As before, we take for simplicity the limit when the observer is in the
infinity. So we get (see Figure 8)
\begin{eqnarray}  \label{row_48}
\frac{\omega_{obs}}{\omega_{s}} = \sqrt{\frac{r^{w}_{s}}{r^{w}_{s} - 1}} \;
\;\; {\rm where} \; \;\; r^{w}_{s} = \frac{r_{s}}{|A|} \; \; .
\end{eqnarray}

We obtained the result that the nearer the source is to the surface given by
the equation $r = |A| $ the more the emitted photon which reaches the
observer is blueshifted.

\section{Appendix~A.2: Scale invariance}

Let us rewrite the Hamilton-Jacobi equation (see Eq.(\ref{row_50})) in the
following way
{\small
\begin{eqnarray}  \label{row_si1}
& & \!\!\!\!\!\!\! \!\!\!\!\!\!\!
\frac{r^{w} + 1}{r^{w}} \left( \frac{\partial S}{\partial t^{w}}
\right)^{2} - \frac{r^{w} + 1}{r^{w}} \: \left( \frac{\partial S}{\partial
r^{w}} \right)^{2} - \frac{1}{(r^{w})^{2}} \: \left( \frac{\partial S}{
\partial \Theta} \right)^{2}  \nonumber \\
& & \!\!\!\!\!\!\! \!\!\!\!\!\!\! - \frac{1}{(r^{w})^{2} \: sin^{2} \Theta} \: \left( \frac{\partial S}{
\partial \Phi} \right)^{2} - \frac{1}{(d^{w})^{2}} \: \frac{r^{w}}{r^{w} + 1}
\: \frac{1}{cos^{2} \vartheta} \: \left( \frac{\partial S}{\partial \vartheta
} \right)^{2}   \nonumber \\
& & \!\!\!\!\!\!\! \!\!\!\!\!\!\! - \frac{1}{(d^{w})^{2}} \: \frac{r^{w}}{r^{w} + 1} \: \left( \frac{
\partial S}{\partial \varsigma} \right)^{2} - ( l^{w} \, c )^{2} = 0 \; ,
\end{eqnarray}
}
where
\begin{eqnarray}  \label{row_si2}
r^{w} = \frac{r}{A} \; , \; \; t^{w} = \frac{ c\;t }{A} \; , \; \; d^{w} =
\frac{d}{A} \; , \; \; l^{w} = m \, A \; .
\end{eqnarray}
From the form of this equation we can notice that the model is explicitly
scale invariant. It means that the systems with different values of $r$, $A$,
$d$, $m$ and $t$ have the same physical properties provided the values of
$l^{w}$, $r^{w}$, $t^{w}$ and $d^{w}$ are the same. In the other words this
means that whenever the scalar field $\varphi$ is present the classical
picture of the world follows the same patterns from the micro to the
marco-scale. The discussion of quantum aspects shall be presented in a
separate paper
but it is worth noting here that the
Klein-Gordon equation for our six-dimensional model possesses similar scaling
properties as the Hamilton-Jacobi equation above.

\section*{Acknowledgments}

This work is supported by L.J.CH.

\newpage

\section{Figure captions}

Figure 1. The potential ${\cal U}_{eff}$ (in units of $m c^{2}$)
(see Eq.(\ref{row_69})) for $k_{\vartheta \varsigma} = 0$ and
different values of ${\cal M}_{\Phi}$ as a function of the
relative radius $r^{w} = r/A$. Horizontal arrow above the curves
denotes the direction of increasing angular momentum ${\cal
M}_{\Phi}.$  The minimum of the potential ${\cal U}_{eff}$
determines the radius $r^{w}$ (and hence the value of $r = r^{w}
\, A$) of a stable circular orbit. The radii $r^{w}$ of stable
orbits correspondes to minima of the potentials depicted on
Figure 1 are $7.5 \; 10^9$, $1.5 \; 10^{10}$ and $3. \; 10^{10}$
respectively. The middle curve (continuous) with $A$ parameter
equal to $5. \, 10^{-7} \; pc$ illustrates the effective potential
for a stable orbit located at a distance equal to the distance of
the Sun from the center of the $\varphi$ scalar field in our
Galaxy
i.e. $r_{\odot} = 7.5 \; kpc.$ \\

Figure 2. The potential ${\cal U}_{eff}$ (in units of $m c^{2}$)
(see Eq.(\ref{row_69})) for different values of $k_{\vartheta
\varsigma}$ and fixed value of ${\cal M}_{\Phi}$ (chosen for
$k_{\vartheta \varsigma} = k_{\vartheta \varsigma,min}(r^{w} =
1/2)$, see Eq.(\ref{lowerlimit})) as a function of the relative
radius $r^{w} = r/A$. The curve for $k_{\vartheta
\varsigma,min}(r^{w} = 1/2)$ (continuous line) has a minimum at
$r^{w}$ equal to $1/2$ and a maximum at finite $r^{w}$ (equal to
$ \approx 2.8 \,$). \\

Figure 3. The proper transversal velocity $v_{\tau}^{st}$ (see
Eq.(\ref{row_83})) of a particle (in units of the velocity of
light c) moving along a stable circular orbit given by
Eqs.(\ref{row_74}),(\ref{row_75}) for different (but fixed for
each curve) values of $k_{\vartheta \varsigma}$ as a function of
the relative radius $r^{w} = r/A$. If $k_{\vartheta
\varsigma}^{2} > - \frac{1}{2} \, m^2 \, c^2 $ then all values of
the radius $r^{w}$ (or $r^{w}_{l} = r_{l}/A $ cf.
Eq.(\ref{row_32})) are allowed for stable orbits. If
$k_{\vartheta \varsigma,min}^{2}(r^{w}_{max}) \leq k_{\vartheta
\varsigma}^{2} < - \frac{1}{2} \, m^2 \, c^2 $ then the stable
orbits, for fixed $k_{\vartheta \varsigma}$ and angular momentum
${\cal M}_{\Phi}$ chosen according to Eq.(\ref{row_77}), may
exist only up to a radius $r_{max}$ (see
Eqs.(\ref{rmax2}),(\ref{rmax1})). (As in Figure 2 $r^{w}_{max} =
1/2$ (then $r^{w}_{0} = 3.5 \,$)). For $m \neq 0$ the curve drawn
in dotted line has physical meaning only up
to $r^{w}_{max} = 1/2.$ \\

Figure 4. The acceleration $a_{r}$ (in units of $1/A$) of a
particle moving along a radial trajectory (see
Eq.(\ref{row_73})). As in Figure 2 $k_{\vartheta \varsigma}$ is
chosen equal to $k_{\vartheta \varsigma,min}(r^{w}_{max} = 1/2)$.
Consequently $a_{r} = 0$ for the relative radius $r^{w} =
r^{w}_{0} = 3.5$ (see Eq.(\ref{rmax1}),(\ref{rmax2})) and the
``particle'' is attracted to the center for all values of $r^{w}
< r^{w}_{0}$ and repelled for $r^{w} > r^{w}_{0}$. For simplicity
we have chosen $r^{w}_{o} = r_{o}/A = r^{w}_{0}$ (see text). For
other curves the particle is attracted to the center
($a_{r} < 0$) for all values of $r^{w}$. \\

Figure 5. Radial acceleration $a_{r}$ of a particle (see
Eq.(\ref{row_73})) for $k_{\vartheta \varsigma} = k_{\vartheta
\varsigma,min}(r_{max} = A/2)$ where $A = 2. \; 10^{5} \, pc$ has
been chosen as typical for quasar-galaxy systems (see Table~1 in
Section~3). \\

Figure 6. Doppler shift (maximal) $z_{D}$ caused by motion of a
particle along a stable circular orbit (see Eq.(\ref{row_z1}))
for different values of internal momentum. The curve $z_{g}$
denotes the gravitational redshift (see Eq.(\ref{row_z2})).
Analogously as in Figure 3 the dotted line has physical
meaning only up to $r^{w}_{max} = 1/2.$ \\

Figure 7. The combined effect of gravitational and Doppler
redshifts (see Eq.(\ref{row_z3})). Analogously as in Figure 3 the
dotted line has physical
meaning only up to $r^{w}_{max} = 1/2.$ \\

Figure 8. The ratio of the frequency $\omega_{obs}$ of the photon
which reaches the observer to the frequency $\omega_{s}$ of the
photon emitted from the source as a function of the relative
distance $r^{w}_{s} = \frac{r_{s}}{|A|}$ of the source from the
center of $\varphi$ field ($A < O$).


\addcontentsline{toc}{section}{Bibliography}

\begin{thebibliography}{11}

\bibitem{Gr-Sch-1981}
M.B. Green, J.H. Schwarz, {\it Nucl.Phys.} {\bf B~181}, 502~(1981).

\bibitem{Gr-Sch-1982b}
M.B. Green, J.H. Schwarz, {\it Nucl.Phys.} {\bf B~198}, 441~(1982).

\bibitem{Kaku-1988}
M. Kaku, "Introduction to Superstrings", Springer, New~York~1988.

\bibitem{Wesson-1992}
P. Wesson, {\it Ap.J} {\bf 394}, 19~(1992).

\bibitem{LORD-Biesiada-Syska-Manka}
..., M.~Biesiada, R.~Ma\'{n}ka, J.~Syska, {\it Inter. Journal of Modern Physics} {\bf D}, Vol. 9 ,Nu.1,~(2000); \\
..., R.~Ma\'{n}ka, J.~Syska, ``Nonhomogeneous six-dimen\-sional Kaluza-Klein compactifi\-ca\-tion'',
pre\-print~U\'{S}L-TP-95/03, University of Silesia,~1995. \\

\bibitem{Biesiada-Rudnicki-Syska-1b}
..., M. Biesiada, K.Rudnicki, J. Syska, "An alternative picture of the structure of galaxies",
in "Gravitation, Electromagnetism and Cosmlogy: toward a new synthesis", ed. K. Rudnicki,
Apeiron, Montreal~2001,~31.

\bibitem{Oort-32}
J.H.~Oort,~{\it Bull.Astr.Inst.Netherlands}~{\bf 6}, \\ 249~(1932).

\bibitem{Rubin-Ford-70}
V.C.Rubin and W.K. Ford, {\it Ap.J} {\bf 379},~(1970).

\bibitem{Rubin-Ford-Thonnard-80}
V.C.Rubin, W.K. Ford and V.C. Rubin, {\it Ap.J} {\bf 471},~(1980).

\bibitem{Gates-Gyuk-Turner}
E. Gates, G. Gyuk and M. Turner, {\it Phys.Rev.Lett.} {\bf 74}, 3724~(1995).

\bibitem{Bahcall-Lubin-Dorman}
N. Bahcall, L.M. Lubin and V. Dorman, {\it Ap.J} {\bf 447}, L81~(1995).

\bibitem{Nishino-Sezgin}
M. Nishino and E. Sezgin, {\it Phys.Lett.} {\bf B~144}, 187~(1984).

\bibitem{Salam-Sezgin-1984}
A. Salam and E. Sezgin, {\it Phys.Lett.} {\bf B~147}, 47~(1984).

\bibitem{Iv-Mel-1995}
V.D. Ivashchuk and V.N. Melnikov, {\it Grav. \& Cosmol.} {\bf 1} No2, 133~(1995); \\
V.D. Ivashchuk, PhD Dissertation, VNICPV, Moscow, 1989.

\bibitem{Br-Mel-1995}
K.A. Bronnikov and V.N. Melnikov, {\it Annals of Physics (N.Y.)} {\bf 239}, 40~(1995).

\bibitem{Iv-Mel-1994}
V.D. Ivashchuk and V.N. Melnikov, {\it Class. Quantum Grav.} {\bf 11}, 1793~(1994).

\bibitem{Manka-Syska-a}
R. Ma\'{n}ka and J. Syska, {\it J.Phys.G: Nucl.Part.Phys.} {\bf 15}, 751-764~(1989).

\bibitem{Land-Lif}
L.D. Landau and E.M. Lifshitz, "The classical Theory of Fields",
Pergamon Press, New~York~1975.

\bibitem{Rubinowicz-Krolikowski-1980}
W. Rubinowicz and W. Kr{\'o}likowski, "Mechanika teoretyczna", Polish
Scientific Publishers, Warszawa~1980. \\
E.T. Whittaker, "A Treatise on the Analytical
Dynamics of Particles and Rigid Bodies", Cambridge Univ. Press, London~1904.

\bibitem{Dziekuje Ci Panie Jezu Chryste}
..., J. Syska, "Self-consistent classical fields in field theories", PhD thesis,
University of Silesia, unpublished,~1995/99.

\bibitem{Hu-1996}
B.L.~Hu,~"General Relativity as Geometro-Hy\-dro\-dy\-na\-mics",
gr-qc/9607070,~(1996).

\bibitem{Smolin-1995}
L. Smolin, "Cosmology as a Problem in Critical Phenomena",
gr-qc/9505022,~(1995).

\bibitem{Tanaka-1995}
Y. Tanaka, et al., {\it Nature} {\bf 375}, 659~(1995).

\bibitem{Arp-1986}
H. Arp, {\it A\&A}~{\bf 156}, 207~(1986).

\bibitem{Tifft-1988}
W.G. Tifft, {\it in} "New Ideas in Astronomy", eds. F.~Bertola,
J.~Sulentic and B.~Madore, Cambridge University Press~1988,~173.

\bibitem{Tifft-93}
W.G. Tifft, Preprint of Steward Observatory No.1143,~(1993).

\bibitem{Arp-Giraud-Sulentic-Vigier-1983}
H. Arp, E. Giraud, J.W. Sulentic, J.P. Vigier,  {\it A\&A}~{\bf 121},
{\bf No.1}, 1-26~(1983).

\bibitem{Burbidge}
G. Burbidge and M. Burbidge, "Quasi-Stellar Objects", Freeman,
San Francisco~1967.\\
H. Arp, E. Giraud, J.W. Sulentic, J.P. Vigier,
{\it A\&A} {\bf 121}, {\bf No.1}, 1,26~(1983).\\
H. Arp, "Quasars, redshifts and controversies", Interstellar Media,
Berkeley~1987. \\
G. Burbidge and A. Hewitt, {\it Sky and Telescope}, 32, December~(1994).

\end{thebibliography}
\end{document}